\documentstyle{article}
\begin{document}

\centerline {{\large\bf PHYSICAL STRUCTURES. }}
\centerline {{\large\bf  FORMING PHYSICAL FIELDS AND MANIFOLDS}}

\centerline {{\large (Properties of skew-symmetric differential forms)}}

\centerline {L.I. Petrova}
\centerline{{\it Moscow State University, Russia, e-mail: ptr@cs.msu.su}}
\bigskip

It is shown that physical fields are formed by physical structures, 
which in their properties are differential-geometrical structures.

These results have been obtained due to using the mathematical apparatus 
of skew-symmetric differential forms. 
This apparatus discloses the controlling role of the conservation 
laws in evolutionary processes, which proceed in material media and 
lead to origination of physical structures and forming physical fields 
and manifolds.

\section{Physical structures} 

The closure conditions of the inexact exterior differential form and 
dual form (the equality to zero of differentials of these forms) can be 
treated as a definition of some differential-geometrical structure. 
In this section it will be shown that as the physical structures, which 
form physical fields, it serve those, which in their properties are 
such differential-geometrical structures. 

The properties of such differential-geometrical structures, 
and correspondingly physical structures, are based on 
the properties of closed exterior differential forms. 

Below the properties  of closed 
exterior differential forms are briefly described. 
(In more detail about skew-symmetric differential forms one can 
read in [1,2]. With the theory of exterior differential forms 
one can become familiar from the works [3-7]). 

\subsection*{Closed exterior differential forms} 

The exterior differential form of degree $p$ ($p$-form on the 
differentiable manifold) is called a closed one if its differential 
equals zero: 
$$
d\theta^p=0\eqno(1) 
$$

From condition (1) one can see that the closed form is a conservative 
quantity. This means that such a form can correspond 
to the conservation law, namely, to some conservative physical quantity. 

If the form is closed on  pseudostructure only 
(i.e. it is the closed {\it inexact} differential form), the closure 
condition is written as 
$$
d_\pi\theta^p=0\eqno(2) 
$$
And the pseudostructure $\pi$ is defined from the condition 
$$
d_\pi{}^*\theta^p=0\eqno(3) 
$$
where ${}^*\theta^p$ is the dual form. 

\{Cohomology, sections of cotangent bundles, the eikonal surfaces, the 
characteristical and potential surfaces, 
and so on can be regarded as examples of pseudostructures. 
For the properties of dual forms see [7].\} 

From conditions (2) and (3) one can see that the exterior differential 
form closed on pseudostructure (a closed inexact form) is a conservative 
object, namely, this quantity conserves on pseudostructure. This can 
also correspond to some conservation law, i.e. to conservative object. 
Such conservation laws that state the existence of 
conservative physical quantities or objects can be named exact ones. 

The pseudostructure and the closed exterior form defined on the 
pseudostructure form a differential-geometrical structure. (It can be 
noted that this structure is the example of the differential-geometrical 
G-Structures). It is evident that just such structures, which correspond 
to the exact conservation law, are physical structures, from which 
physical fields are formed. 
\{The physical fields [6] are a special form of the substance, 
they are carriers of various interactions such as electromagnetic, 
gravitational, wave, nuclear and other kinds of interactions.\} 

The problem of how these structures arise and how 
physical fields are formed will be discussed below. 

Thus, the closure conditions for the exterior differential 
form ($d_{\pi }\,\theta ^p\,=\,0$) 
and the dual form ($d_{\pi }\,^*\theta ^p\,=\,0$) are 
mathematical expressions of the exact conservation law and they 
define the physical structure. 

The mathematical expression for the exact conservation law and 
its connection with physical fields can be schematically written 
in the following way 
$$
\def\\{\vphantom{d_\pi}}
\cases{d_\pi \theta^p=0\cr d_\pi {}^{*\mskip-2mu}\theta^p=0\cr}\quad
\mapsto\quad
\cases{\\\theta^p\cr \\{}^{*\mskip-2mu}\theta^p\cr}\quad\hbox{---}\quad
\hbox{physical structures}\quad\mapsto\quad\hbox{physical fields}
$$

Since the relations for exact conservation laws and for relevant 
physical structures (that form physical fields) are expressed in 
terms of the closed exterior and dual forms, it is evident the 
field theories (which describe physical fields) are based on the 
mathematical apparatus of the closed exterior differential and 
dual forms. 

One can express the field theory operators in terms of following 
operators of exterior differential forms: $d$ (exterior differential), 
$\delta$ (the operator of transforming the form of degree $p+1$ into the 
form of degree $p$), $\delta '$ (the operator of cotangent transforms), 
$\Delta $ (that of the transformation $d\delta-\delta d$), $\Delta '$ (the 
operator of the transformation $d\delta'-\delta' d$). In terms of these 
operators, which act onto exterior forms, one can write down the 
operators by Green, d'Alembert, Laplace and the operator of canonical 
transform [7]. 

The equations, that are equations of the existing field theories, are
those obtained on the basis of the properties of the exterior 
differential form theory. 

It can be shown that to the quantum mechanical equations (to the 
equations by Shr\H{o}dinger, Heisengerg and Dirac) there correspond 
the closed exterior forms of zero degree or the relevant dual forms. 
The closed exterior form of zero degree corresponds to the 
Schr\H{o}dinger equation, the close dual form corresponds to the 
Heisenberg equation. It can be pointed out that, whereas the equations 
by Schr\H{o}dinger and Heisenberg describe a behavior of the potential 
obtained from the zero degree closed form, Dirac's {\it bra-} and 
{\it cket}- vectors constitute a zero degree closed exterior form 
itself as the result of conjugacy (vanishing the scalar product). 

The Hamilton formalism is based on the properties of closed 
exterior and dual forms of the first degree. The closed exterior 
differential form $ds=-Hdt+p_j dq_j$ (the Poincare invariant) 
corresponds to the field equation. 

The properties of closed exterior and dual forms of the second degree 
lie at the basis of the electromagnetic field equations. The Maxwell 
equations may be written as  $d\theta^2=0$, $d^*\theta^2=0$ [7], where 
$\theta^2=\frac{1}{2}F_{\mu\nu}dx^\mu dx^\nu$ (here $F_{\mu\nu}$ is 
the strength tensor). 

Closed exterior and dual forms of the third degree correspond to the 
gravitational field. 

From the above stated one can see that to each type of physical fields 
there corresponds a closed exterior form of appropriate degree. 

The connection of the physical structures with exterior forms 
allows to understand the properties and specific features of the 
physical structures.

\section{The properties of exterior differential forms and 
physical structures} 

Basic properties of exterior differential forms are connected with 
the fact that any closed form is a differential. 

The exact form is, by definition, a differential 
$$
\theta^p=d\theta^{p-1}\eqno(4)
$$
In this case the differential is total. The closed inexact form is 
a differential too. The closed inexact form is an interior (on 
pseudostructure) differential, that is 
$$
\theta^p_\pi=d_\pi\theta^{p-1}\eqno(5) 
$$

And so, any closed form is a differential of the form of a lower 
degree: the total one $\theta^p=d\theta^{p-1}$ if the form is exact, 
or the interior one $\theta^p=d_\pi\theta^{p-1}$ on pseudostructure if 
the form is inexact. (This may have the physical meaning: the form of 
lower degree can correspond to the potential, and the closed form by 
itself can correspond to the potential force.)

\subsection*{Invariant properties of closed exterior differential forms 
and physical structures. Nondegenerate transformations} 

Since the closed form is a differential, then it is evident 
that the closed form proves to be invariant under all 
transformations that conserve a differential. 

The examples of such nondegenerate transformations are unitary, tangent, 
canonical, and gradient transformations. 

To the nondegenerate transformations there are assigned closed forms 
of given degree. To the unitary transformations it is assigned (0-form), 
to the tangent and canonical transformations it is assigned (1-form), 
to the gradient transformations it is assigned (2-form) and so on. 
It should be noted that these transformations are {\it gauge 
transformations} for spinor, scalar, vector, tensor (3-form) fields. 
Hence one can see that the physical structure relate to the gauge type 
of the differential-geometrical G-Structure. They remain to be 
invariant under all transformations that conserve the differential.

It is well known that these are transformations typical for existing 
field theories. The equations of existing field theories remain 
invariant under such transformations. 

The closure of exterior differential forms, and hence their invariance, 
results from the conjugacy of elements of exterior or dual forms. 

From the definition of exterior differential form one can see that 
exterior differential forms have complex structure. Specific features 
of the exterior form structure are homogeneity with respect to the 
basis, skew-symmetry, integrating terms each including two objects of 
different nature (the algebraic nature for form coefficients, and the 
geometric nature for  base components). Besides,  the exterior form 
depends on the space dimension and on the manifold topology. The 
closure property of the exterior form means that any objects, namely, 
elements of the exterior form, components of elements, elements of 
the form differential, exterior and dual forms and others, turn out to 
be conjugated. (In the author's work [8] some types of conjugacy of 
exterior differential forms have been considered). A variety 
of objects of conjugacy leads to the fact that the closed forms can 
describe a great number of various physical structures. 

Since the conjugacy is a certain connection between two operators or 
mathematical objects, it is evident that, to express a conjugacy 
mathematically, it can be used relations. Just such relations constitute 
the basis of mathematical apparatus of the exterior differential forms. 
This is an identical relation. Identical relations of exterior 
differential forms also disclose the properties of 
physical structures. 

\subsection*{Identical relations of exterior differential forms} 

Identical relations of exterior differential forms reflect the 
closure conditions of differential forms, namely, vanishing the form 
differential (see formulas (1), (2), (3)) and hence 
the conditions connecting the forms of consequent degrees (see formulas 
(4), (5)). Since the closure conditions of differential forms 
and dual forms specify the differential-geometrical structures and 
physical structures, identical relations for exterior differential 
forms specify the physical structures. 

The identical relations are a mathematical expression of the invariance 
and covariance. And this lies at the basis of existing field theories. 

Examples of such relations are canonical relations in the Schr\H{o}dinger 
equations, gauge invariance in electromagnetic theory, commutator 
relations in the Heisenberg theory, symmetric connectednesses, identity 
relations by Bianchi in the Einstein theory, cotangent bundles in 
the Yang-Mills theory, the covariance conditions in the tensor methods, 
the characteristic relations (integrability conditions) in equations 
of mathematical physics, etc. (In more detail about identical (and 
nonidentical) relations it is outlined in the author's work [8]). 

The identical relations express the fact that each closed exterior 
form is a differential of some exterior form (with a degree less 
by one). In general form such an identical relation can be written as 
$$
d _{\pi}\phi=\theta _{\pi}^p\eqno(6)
$$
In this relation the form in the right-hand side has to be a 
{\it closed} one. (As it will be shown below, 
the identical relations are satisfied only on pseudostructures). 

In identical relation (6) in one side it stands the closed form and 
in other side does a differential of some differential 
form of the less by one degree, which is a closed form as well. 

The identical relations of another type are the analog of relation (6) 
obtained by differentiating or integrating this relation.

\section{Mechanism of origination of physical structures} 

It has been shown that the skew-symmetric closed exterior differential 
forms allow to describe the properties and specific features of the 
physical structures. 

In this section it will be shown that the skew-symmetric 
differential forms describe also the process of origination of 
physical structures. However, to do this, one must 
use skew-symmetric differential forms, which, in contrast to 
exterior (skew-symmetric) differential forms, possess the 
evolutionary properties, and for this reason they were named 
evolutionary differential forms. 

A peculiarity of the evolutionary differential forms consists in the 
fact that they generate exterior differential forms, which correspond 
to physical structures. This elucidates the process of origination of 
physical structures.

\subsection*{Specific features of the evolutionary differential forms} 

A radical distinction between the evolutionary forms and the exterior 
ones consists in the fact that the exterior differential forms are 
defined on manifolds with {\it closed metric forms}, whereas the 
evolutionary differential forms are defined on manifolds with {\it 
unclosed metric forms}. 

The closed metric forms define the manifold structure, and the 
commutators of metric forms define the manifold differential 
characteristics that specify the manifold deformation: bending, 
torsion, rotation, twist. 

The theory of exterior differential forms was developed for both  
differentiable manifolds and manifolds with structures of any types as 
well. It is evident that the manifolds, which are metric ones or possess 
the structure, have closed metric forms. All have one common property, 
namely, locally they admit one-to-one mapping into  Euclidean subspaces 
and into other manifolds or submanifolds of the same dimension [5]. 

When describing any processes in terms of differential equations, 
one has to deal with manifolds that do not allow one-to-one 
mapping described above. Such manifolds are, for example, manifolds 
formed by trajectories of elements of the system described by 
differential equations. The manifolds that can be called accompanying 
manifolds are variable deforming manifolds. 

If the manifolds are deforming manifolds, this means that their 
metric form commutators are nonzero. That is, the metric forms of 
such manifolds turn out to be unclosed. The accompanying manifolds 
and manifolds occurring to be deforming ones are examples of such 
manifolds. 

The skew-symmetric evolutionary differential forms, whose basis are 
deforming manifolds, are defined on manifolds with unclosed metric forms. 

The evolutionary properties of the evolutionary skew-symmetric 
differential forms are just connected with properties of the metric 
form commutators. 

\bigskip 
The evolutionary differential form of degree $p$ ($p$-form) 
can be also written as an exterior differential form [1]. 

But the evolutionary form differential cannot be written similarly to 
that presented for exterior differential forms. In the evolutionary form 
differential there appears an additional term connected with 
the fact that the basis of the form changes [1]. 

For example, we again inspect the first-degree form 
$\omega=a_\alpha dx^\alpha$. 
[From here on the symbol $\sum$ will be omitted and it will be 
implied that a summation over double indices is performed.  Besides, 
the symbol of exterior multiplication will be also omitted for the 
sake of presentation convenience]. 

The differential of this form can 
be written as $d\omega=K_{\alpha\beta}dx^\alpha dx^\beta$, where 
$K_{\alpha\beta}=a_{\beta;\alpha}-a_{\alpha;\beta}$ are 
components of the commutator of the form $\omega$, and 
$a_{\beta;\alpha}$, $a_{\alpha;\beta}$ are the covariant 
derivatives. If we express the covariant derivatives in terms of 
the connectedness (if it is possible), they can be written 
as $a_{\beta;\alpha}=\partial a_\beta/\partial 
x^\alpha+\Gamma^\sigma_{\beta\alpha}a_\sigma$, where the first 
term results from differentiating the form coefficients, and the 
second term results from differentiating the basis. (In 
Euclidean space covariant derivatives coincide with ordinary ones 
since in this case derivatives of the basis vanish). If 
we substitute the expressions for covariant derivatives into the 
formula for the commutator components, we obtain the following 
expression for the commutator components of the form $\omega$: 
$$
K_{\alpha\beta}=\left(\frac{\partial a_\beta}{\partial
x^\alpha}-\frac{\partial a_\alpha}{\partial
x^\beta}\right)+(\Gamma^\sigma_{\beta\alpha}-
\Gamma^\sigma_{\alpha\beta})a_\sigma\eqno(7)
$$
Here the expressions 
$(\Gamma^\sigma_{\beta\alpha}-\Gamma^\sigma_{\alpha\beta})$ 
entered into the second term are just the components of 
commutator of the first-degree metric form. 

The evolutionary form commutator of any degree involves the commutator 
of the manifold metric form of corresponding degree. The commutator of 
the exterior form does not contains a similar term because the 
commutator of metric form of manifold, on which the exterior form is 
defined, is equal to zero. 

The commutators of evolutionary forms depend not only on the 
evolutionary form coefficients, but also on the characteristics of 
manifolds, on which this form is defined. As a result, such a dependence 
of the evolutionary form commutator produces the topological 
and evolutionary properties of both the commutator and the evolutionary 
form itself (this will be demonstrated below). 

The evolutionary differential form commutator, in contrast to that of 
the exterior one, cannot be equal to zero since it includes the metric 
form commutator being nonzero. This means that the evolutionary form 
differential is nonzero. Hence, {\it the evolutionary differential form, 
in contrast to the case of the exterior form, cannot be closed}. 

It was noted above that the closed exterior differential forms describe 
the conservation laws. It appears that the evolutionary differential 
forms, which are unclosed, describe the conservation laws also [9]. The 
difference consists in the fact that the closed exterior differential 
forms describe the conservation laws for physical fields (exact 
conservation laws that state the existence of conservative physical 
quantities or objects), whereas the evolutionary differential forms 
describe the conservation laws for material media (material systems). 
\{The material system is a variety of elements that have internal 
structure and interact to one another. As examples of material systems 
it may be thermodynamic, gas dynamical, cosmic systems, systems of 
elementary particles (pointed above) and others. Any material media are 
such material systems. Examples of elements that constitute the material 
system are electrons, protons, neutrons, atoms, fluid particles, cosmic 
objects and others\}. 

The conservation laws for material systems establish the balance 
between the variation of a physical quantity and the corresponding 
external action. These conservation laws, which can be called the 
balance conservation laws, are the conservation laws for energy, linear 
momentum, angular momentum, and mass. 

The properties of evolutionary differential forms, which correspond to 
the balance conservation laws, lie at the basis of the 
processes of forming physical fields and corresponding manifolds. 
To understand a mechanism of these processes one should look at the 
properties of the balance conservation laws.

\subsection*{Properties of the balance conservation laws} 

Equations of the balance conservation laws are differential 
(or integral) equations that describe a variation 
of functions corresponding to physical quantities [10-13]. 
From the equations, which describe the balance conservation laws, 
the evolutionary relation in differential forms is obtained. 

Let us analyze the equations that describe the balance conservation 
laws for energy and linear momentum. 

We introduce two frames of reference: the first is an inertial one (this 
frame of reference is not connected with the material system), and 
the second is an accompanying one (this system is connected 
with the manifold constructed of the trajectories of 
the material system elements). The energy equation 
in the inertial frame of reference can be reduced to the form: 
$$
\frac{D\psi}{Dt}=A_1 \eqno(8)
$$
where $D/Dt$ is the total derivative with respect to time, $\psi $ 
is the functional of the state that specifies the material system, 
$A_1$ is the quantity that depends on specific features of the system 
and on external energy actions onto the system. \{The action functional, 
entropy, wave function can be regarded as examples of the functional 
$\psi $. Thus, the equation for energy presented in terms of the action 
functional $S$ has a similar form: $DS/Dt\,=\,L$, where $\psi \,=\,S$, 
$A_1\,=\,L$ is the Lagrange function. In mechanics of continuous media  
the equation for energy of ideal gas can be presented in the form [11]: 
$Ds/Dt\,=\,0$, where $s$ is entropy. In this case $\psi \,=\,s$, 
$A_1\,=\,0$. It is worth noting that the examples presented show that 
the action functional and entropy play the same role.\} 

In the accompanying frame of reference the total derivative with respect 
to time is transformed into the derivative along the trajectory. 
Equation (8) is now written in the form 
$$
{{\partial \psi }\over {\partial \xi ^1}}\,=\,A_1 \eqno(9)
$$
here $\xi^1$ is the coordinate along the trajectory. 

In a similar manner, in the accompanying frame of reference the equation 
for linear momentum turns out to be reduced to the equation of the form 
$$
{{\partial \psi}\over {\partial \xi^{\nu }}}\,=\,A_{\nu },\quad \nu \,=\,2,\,...\eqno(10)
$$
where $\xi ^{\nu }$ are the coordinates in the direction normal to the 
trajectory, $A_{\nu }$ are the quantities that depend on the specific 
features of the system and external force actions. 

Eqs. (9), (10) can be convoluted into the relation 
$$
d\psi\,=\,A_{\mu }\,d\xi ^{\mu },\quad (\mu\,=\,1,\,\nu )\eqno(11) 
$$
where $d\psi $ is the differential expression 
$d\psi\,=\,(\partial \psi /\partial \xi ^{\mu })d\xi ^{\mu }$. 

Relation (11) can be written as 
$$
d\psi \,=\,\omega \eqno(12)
$$
Here $\omega \,=\,A_{\mu }\,d\xi ^{\mu }$ is the differential form 
of the first degree. 

Since the balance conservation laws are evolutionary ones, the relation 
obtained is also an evolutionary relation. 

Relation (12) was obtained from the equation of the balance 
conservation laws for energy and linear momentum. In this relation 
the form $\omega $ is that of the first degree. If the equations of the 
balance conservation laws for angular momentum be added to the equations 
for energy and linear momentum, this form in the evolutionary relation 
will be the form of the second degree. And in  combination with the 
equation of the balance conservation law of mass this form will be the 
form of degree 3. 

Thus, in the general case the evolutionary relation can be written as 
$$
d\psi \,=\,\omega^p \eqno(13)
$$
where the form degree  $p$ takes the values $p\,=\,0,1,2,3$. 
(The evolutionary relation for $p\,=\,0$ is similar to that in the 
differential forms, and it was obtained from the interaction of energy 
and time.) 

In relation (12) the form $\psi$ is the form of zero degree. And in 
relation (13) the form $\psi$ is the form of $(p-1)$ degree. 

Let us show that {\it the evolutionary relation  obtained from the 
equation of the balance conservation laws turns out to be nonidentical}. 

To do so we shall analyze relation (12). 

Let us consider the commutator of the form 
$\omega \,=\,A_{\mu }d\xi ^{\mu }$. The components of the commutator 
of such a form (as it was pointed above) can be written as follows:
$$
K_{\alpha \beta }\,=\,\left ({{\partial A_{\beta }}\over {\partial \xi ^{\alpha }}}\,-\,
{{\partial A_{\alpha }}\over {\partial \xi ^{\beta }}}\right )\eqno(14) 
$$
(here the term  connected with the nondifferentiability of the manifold 
has not yet been taken into account). 

The coefficients $A_{\mu }$ of the form $\omega $ have been obtained 
either from the equation of the balance conservation law for energy or 
from that for linear momentum. This means that in the first case the 
coefficients depend on the energetic action and in the second case they 
depend on the force action. In actual processes energetic and force 
actions have different nature and appear to be inconsistent. The 
commutator of the form $\omega $ constructed of the derivatives of such 
coefficients is nonzero. This means that the differential of the form 
$\omega $ is nonzero as well. Thus, the form $\omega$ proves to be 
unclosed. 

This means that the evolutionary relation cannot be an identical one. 
In the left-hand side of this relation it stands a differential, 
whereas in the right-hand side it stands an unclosed form, which is not 
a differential. 

A role of the evolutionary relation obtained consists in the following. 
It enables one to describe the evolutionary processes in material medium, 
which lead to emergence of physical structures that form physical fields. 
From the nonidentical evolutionary relation the identical relations, 
which contain closed exterior forms, are obtained (under the degenerate 
transformations). The emergence of the closed exterior form points to 
a rise to the physical structure. 

Here it should be noted that the evolutionary nonidentical relation is 
a selfvarying relation. This plays a governing role while describing 
the evolutionary processes.

\subsection*{Selfvariation of the evolutionary nonidentical relation} 

The evolutionary nonidentical relation is selfvarying, 
because, firstly, it is nonidentical, namely, it contains 
two objects one of which appears to be unmeasurable, and, 
secondly, it is an evolutionary relation, namely, a variation of 
any object of the relation in some process leads to variation of 
another object and, in turn, a variation of the latter leads to 
variation of the former. Since one of the objects is an unmeasurable 
quantity, the other cannot be compared with the first one, and hence, 
the process of mutual variation cannot terminate. 

Varying the evolutionary form coefficients leads to varying the first 
term of the evolutionary form commutator (see (7),(14)). In accordance 
with this variation it varies the second term, that is, the metric form 
of manifold varies. Since the metric form commutators specifies the
manifold differential characteristics, which are connected with the 
manifold deformation (as it has been pointed out, the commutator of the 
zero degree metric form specifies the bend, that of second degree 
specifies various types of rotation, that of the third degree specifies 
the curvature), this points to the manifold deformation. This means that 
it varies the evolutionary form basis. In turn, this leads to variation 
of the evolutionary form, and the process of intervariation of the 
evolutionary form and the basis is repeated. Processes of variation of 
the evolutionary form and the basis are controlled by the evolutionary 
form commutator and it is realized according to the evolutionary relation. 

A significance of the evolutionary relation selfvariation consists in 
the fact that in such a process it can be realized conditions under 
which the identical relation is obtained from the nonidentical relation. 
These are conditions of degenerate transformation.

\subsection*{Obtaining an identical relation from a nonidentical one} 

To obtain the identical relation from the evolutionary nonidentical 
relation, it is necessary that a closed exterior differential form 
would be derived from the evolutionary differential form, which is 
included into evolutionary  relation. However, as it has been shown 
above, the evolutionary form cannot be a closed form. For this reason 
the transition from the evolutionary form is possible only to an 
{\it inexact} closed exterior form, which is defined on pseudostructure. 

To the pseudostructure it is assigned a closed dual form 
(whose differential vanishes). For this reason the transition 
from the evolutionary form to a closed inexact exterior form proceeds 
only when the conditions of vanishing the dual form differential are 
realized, in other words, when the metric form differential or 
commutator becomes equal to zero. 

Since the evolutionary form differential is nonzero, whereas the closed 
exterior form differential is zero, a transition from the evolutionary 
form to the closed exterior form is allowed only under {\it degenerate 
transformation}. The conditions of vanishing the dual form differential 
(the additional condition) are the conditions of degenerate 
transformation. 

Such conditions can just be realized under selfvariation of the 
nonidentical evolutionary relation. 

As it has been already mentioned, the evolutionary differential form 
$\omega^p$, involved into nonidentical relation (13), is an unclosed 
one. The commutator, and hence the differential, of this form is nonzero. 
That is, 
$$
d\omega^p\ne 0\eqno(15)
$$
If the conditions of degenerate transformation are realized, then from 
the unclosed evolutionary form one can obtain a differential form closed 
on pseudostructure. The differential of this form equals zero. That is, 
it is realized the transition 

$d\omega^p\ne 0 \to $ (degenerate transformation) $\to d_\pi \omega^p=0$,
$d_\pi{}^*\omega^p=0$

On the pseudostructure $\pi$ evolutionary relation (13) transforms into 
the relation 
$$
d_\pi\psi=\omega_\pi^p\eqno(16)
$$
which proves to be an identical relation. Indeed, since the form 
$\omega_\pi^p$ is a closed one, on the pseudostructure this form turns 
out to be a differential of some differential form. In other words, 
this form can be written as $\omega_\pi^p=d_\pi\theta$. Relation (16) 
is now written as 
$$
d_\pi\psi=d_\pi\theta
$$
There are differentials in the left-hand and right-hand sides of 
this relation. This means that the relation is an identical one. 

From evolutionary relation (13) it is obtained the identical on the 
pseudostructure relation. Here it should be noted that in this case 
the evolutionary relation itself remains to be nonidentical one. 
(At this point it should be emphasized that differential, which equals 
zero, is an interior one. The evolutionary form 
commutator becomes zero only on the pseudostructure. The total 
evolutionary form commutator is nonzero. That is, under degenerate 
transformation the evolutionary form differential vanishes only 
{\it on pseudostructure}. The total differential of the evolutionary 
form is nonzero. The evolutionary form remains to be unclosed.) 

The transition from nonidentical relation (13) to identical relation 
(16) means the following. Firstly, it is from such a relation that 
one can obtain the differential $d_\pi\psi$ and find the desired 
function $\psi_\pi$ (a potential). And, secondly, an emergence 
of the closed (on pseudostructure) inexact exterior form $\omega_\pi^p$ 
(right-hand side of relation (16)) points to an origination of the 
conservative object. This object is a conservative quantity (the closed 
exterior form  $\omega_\pi^p$) on the pseudostructure (the dual form 
$^*\omega^p$, which defines the pseudostructure). This object is an 
example of the physical structure. 

As conditions of degenerate transformation (additional conditions) 
it can serve any symmetries of the evolutionary form coefficients 
or its commutator. While describing material system such additional 
conditions are related, for example, to degrees of freedom of the 
material system. 

Mathematically to the conditions of degenerate transformation it is 
assigned the requirement that some functional expressions become equal 
to zero. Such functional expressions are Jacobians, determinants, 
the Poisson brackets, residues, and others. 

The degenerate transformation is realized as a transition to 
nonequivalent coordinate system: a transition from the accompanying 
noninertial coordinate system to the locally inertial that. 
Evolutionary relation (13) and condition (15) relate 
to the system being tied to the accompanying manifold, whereas 
identical relations (16) can relate only to the locally inertial 
coordinate system being tied to pseudostructure. 

Thus, while selfvariation of the evolutionary nonidentical 
relation the dual form commutator can vanish (due to the  symmetries of 
the evolutionary form coefficients or its commutator). This means that 
it is formed the pseudostructure on which the differential form turns 
out to be closed. The emergence of the form being closed on 
pseudostructure points out to origination of the physical structures.

\subsection*{Causality of origination of physical structures and of 
forming physical fields} 

Since closed inexact exterior forms corresponding to physical structure 
are obtained from the evolutionary relation for the material system, it 
follows that physical structures are generated by the material systems. 
(This is controlled by the conservation laws.) 

The mechanism of this process involves the following steps. 

1) The external actions onto the material system are transformed into 
the unmeasurable quantity that acts as an internal force and brings the 
material system into the nonequilibrium state. (A nonzero value of 
the evolutionary form commutator. Nonidentical evolutionary relation 
obtained from balance conservation laws). 

2) Selfvariation of the nonequilibrium state of the material system. 
The deformation of accompanying manifold. (Selvariation of the 
nonidentical evolutionary relation. The topological properties of the 
evolutionary form commutator.) 

3) Realization of the degrees of freedom of the material system in the 
process of selvariation of the nonequilibrium state of the system 
itself. (Degenerate transformations). 

4) Transition of the material system from the nonequilibrium state 
into the locally equilibrium one: transition of an internal force 
into a potential force. The emergence of the physical structures. 
(Formation  of the closed (on pseudostructure) inexact exterior form 
and obtaining the state differential, which specifies a state of 
material system). 

In material system the origination of a physical structure reveals 
as a new measurable and observable formation that spontaneously arises 
in the material system. \{{\it As the examples it can be fluctuations, 
pulsations, waves, vortices, creating massless particles.}\}. 

In the physical process this formation is spontaneously extracted from 
the local domain of the material system and so it allows the local 
domain of material system to get rid of an internal force and 
come into a locally equilibrium state. The formation created in a 
local domain of the material system (at the cost of an unmeasurable 
quantity that acts in the local domain as an internal force) and 
liberated from that, begins acting onto the neighboring local domain 
as a force. This is a potential force, this fact is indicated by the 
double meaning of the closed exterior form (on one hand, a 
conservative quantity, and, on other hand, a potential force). 
(This action was produced by the material system in itself, 
and therefore this is a potential action rather than an arbitrary one). 
The transition of the material system from nonequilibrium into 
a locally equilibrium state (which is indicated by the formation of 
a closed form) means that the unmeasurable quantity described by the 
nonzero commutator of the nonintegrable differential form $\omega^p$, 
that acts as an internal force, transforms into the measurable quantity. 
It is evident that this is just the measurable quantity, which acts as a 
potential force. In other words, the internal force transforms into a 
potential force. 
The neighboring domain of the material system works over this action, 
which appears to be external with respect to that. If in the process the 
conditions of conjugacy of the balance conservation laws turn out to be 
satisfied again, the neighboring domain will create a formation by its 
own, and this formation will be extracted from this domain. In such a 
way the formation can move relative to the material system. (Waves are 
the example of such motions). 

The physical structures are generated by numerous local domains of 
the material system and at numerous instants of realizing various 
degrees of freedom of the material system. It is evident that they can 
generate fields. In this manner physical fields are formed. To obtain 
the physical structures that form a given physical field, one has to 
analyze the material system  corresponding to this field and the 
appropriate evolutionary relation. In particular, to obtain the 
thermodynamic structures (fluctuations, phase transitions, etc), one 
has to analyze the evolutionary relation for the thermodynamic systems, 
to obtain the gas dynamic ones (waves, jumps, vortices, pulsations) one 
has to employ the evolutionary relation for gas dynamic systems, for 
the electromagnetic field one must employ a relation obtained from 
equations for charged particles. 

\subsection*{Characteristics of physical structure} 

Since the closed exterior differential form, which corresponds to the 
physical structure arisen, was obtained from the nonidentical relation 
that involves the evolutionary form, it is evident that the physical 
structure characteristics must be connected with those of the 
evolutionary form and of the manifold on which this form is defined, 
with the conditions of degenerate transformation and with the values of 
commutators of the evolutionary form and the manifold metric form. 

The conditions of degenerate transformation, as it was said before, 
determine the pseudostructures. The first term of the evolutionary form 
commutator determines the value of the discrete change (the quantum), 
which the quantity conserved on the pseudostructure undergoes when 
transition from one pseudostructure to another. The second term of the 
evolutionary form commutator specifies the characteristics that fixes 
the character of the initial manifold deformation, which took place 
before the physical structure arose. (Spin is such an example). 

A discrete (quantum) change of a quantity proceeds in the direction 
that is normal (more exactly, transverse) to the pseudostructure. Jumps 
of the derivatives normal to the potential surfaces are examples of 
such changes.

\subsection*{Classification of physical structures} 

Closed forms that correspond to physical structures are generated by the 
evolutionary relation having the parameter $p$, which defines a number 
of interacting balance conservation laws. Therefore, the 
physical structures can be classified by the parameter $p$. 

The other parameter is a degree of closed forms generated by 
the evolutionary relation. 

To determine this parameter one has to consider the problem of 
integration of the nonidentical evolutionary relation. 

Under degenerate transformation from the nonidentical evolutionary 
relation one obtains a relation being identical on pseudostructure. 
Since the right-hand side of such a relation can be expressed in terms 
of differential (as well as the left-hand side), one obtains a relation 
that can be integrated, and as a result he obtains a relation with the 
differential forms of less by one degree. The relation obtained after 
integration proves to be nonidentical as well. 
The resulting nonidentical relation of degree $(p-1)$ (relation that 
includes the forms of the degree $(p-1)$) can be integrated once again 
if the corresponding degenerate transformation has been realized and 
the identical relation has been formed. By sequential integrating 
the evolutionary relation of degree $p$ (in the case of realization of 
the corresponding degenerate transformations and forming the 
identical relation), one can get closed (on the pseudostructure) 
exterior forms of degree $k$, where $k$ ranges from $p$ to $0$. 

Thus, one can see that physical structures, to which there are assigned 
the closed (on the pseudostructure) exterior forms, can depend on two 
parameters. These parameters are the degree of evolutionary form $p$ 
(in the evolutionary relation) and the degree of created closed 
forms $k$. 

In addition to these parameters, another parameter appears, namely, the 
dimension of space. If the evolutionary relation generates the closed 
forms of degrees $k=p$, $k=p-1$, \dots, $k=0$, to them there are 
assigned the pseudostructures of dimensions $(N-k)$, where $N$ is the 
space dimension. \{It is known that to the closed exterior differential 
forms of degree $k$ there are assigned skew-symmetric tensors of rank 
$k$ and to corresponding dual forms there do the pseudotensors of rank 
$(N-k)$, where $N$ is the space dimensionality. The pseudostructures 
correspond to such tensors, but only on the space formed.\}

\subsection*{Forming of pseudometrical manifolds and metric spaces and 
physical fields} 

As mentioned before, the additional conditions, namely, the conditions 
of degenerate transformation, specify the pseudostructure. But at every 
stage of the evolutionary process it is realized only one element of 
pseudostructure, namely, a certain minipseudostructure. 

While varying the evolutionary variable the minipseudostructures form 
the pseudostructure. The example of minipseudoctructure is the wave 
front. The wave front is an eikonal surface (the level surface), i.e. 
the surface with a conservative quantity. 

Manifolds with closed metric forms are formed by pseudostructures. They 
are obtained from manifolds with unclosed metric forms. In this case the 
initial (accompanying) manifold (on which the evolutionary form is 
defined) and the manifold with closed metric forms originated (on which 
the closed exterior form is defined) are different spatial objects. 

It takes place a transition from the initial (accompanying) manifold 
with unclosed metric form to the pseudostructure, namely, to the 
manifold with closed metric forms being created. Mathematically this 
transition (degenerate transformation) proceeds as {\it a transition 
from one frame of reference to another, nonequivalent, frame of 
reference.} 

The pseudostructures, on which the closed {\it inexact} forms are 
defined, form the pseudomanifolds. (Integral surfaces, pseudo-Riemann 
and pseudo-Euclidean spaces are the examples of such manifolds). In this 
process the dimensions of the  manifolds formed  are connected with the 
evolutionary form degree. 

To the transition from pseudomanifolds to metric space it is assigned 
a transition from closed {\it inexact} differential forms to {\it exact} 
exterior differential forms. 

It was shown above that the evolutionary relation of degree $p$ can 
generate (in the presence  of the degenerate transformations) closed 
forms of the degree $0,...,p$.  While generating closed forms of 
sequential degrees  $k=p, k=p-1,..., k=0$ the pseudostructures of 
dimensions $(n+1-k)$ are obtained. As a result of transition to the 
exact closed form of zero degree the metric structure of the dimension 
$n+1$ is obtained. Under the influence of an external action (and with 
the availability of degrees of freedom) the material system can transfer 
the initial inertial space into the space of the dimension $n+1$. 

Sections of the cotangent bundles (Yang-Mills fields), cohomologies by 
de Rham, singular cohomologies, pseudo-Riemannian and pseudo-Euclidean 
spaces, and others are examples of the psedustructures and spaces that 
are formed by pseudostructures. Euclidean and Riemannian spaces are 
examples of metric manifolds that are obtained when going to the exact 
forms. Here it should to be noted that the examples of pseudometric 
spaces are potential surfaces (surfaces of a simple layer, a double 
layer and so on). In these cases the type of potential surfaces is 
connected with the above listed parameters. 

What can be said about the pseudo-Riemannian manifold and Riemannian 
space? 

The distinctive property of the Riemannian manifold is an availability 
of the curvature. This means that the metric form commutator of the 
third degree is nonzero. Hence, it does not equal zero the evolutionary 
form commutator of the third degree, which involves into itself the 
metric form commutator. That is, the evolutionary form, which enters 
into the evolutionary relation, is unclosed, and the relation is 
nonidentical. When realizing pseugostructures of the dimensions 
$1,2,3,4$ and obtaining the closed inexact forms of the degrees 
$k=3, k=2, k=3, k=4$ the pseudo-Riemannian space is formed, and the 
transition to the exact form of zero degree corresponds to the 
transition to the Riemannian space. 

It is well known that while obtaining the Einstein equations it was 
suggested that there are fulfilled the conditions: The Bianchi identity 
is satisfied, the coefficients of connectedness are symmetric, the 
condition that the coefficients of connectedness are the Christoffel 
symbols, and an existence of the transformation, under which the 
coefficients of connectedness vanish. These conditions are the 
conditions of realization of the degenerate transformations for 
nonidentical relations obtained from the evolutionary relation of 
the degree $p=3$  and after going to the exact relations. In this case 
to the Einstein equation it is assigned the identical relation of the 
first degree. 

As it has been pointed above, from the physical structures, to which the 
closed forms are assigned and which are generated by material systems, 
the physical fields are formed. 
Physical fields form the unified whole with corresponding manifolds. 
Since the closed metric form is dual with respect to some closed 
exterior differential form, the metric forms cannot become closed 
by themselves, independent of the closed exterior differential form. 
This proves that manifolds with closed metric forms are connected with 
the closed exterior differential forms. This indicates that the fields 
of conservative quantities, such as physical fields, are formed from 
closed exterior forms at the same instant of time when the manifolds 
are created from the pseudoctructures. (The specific feature of the 
manifolds with closed metric forms that have been formed 
is that they can carry some information.)

1. Petrova L.~I., Exterior and evolutionary skew-symmetric differential 
forms and their role in mathematical physics. 

http://arXiv.org/pdf/math-ph/0310050

2. Petrova L.~I., Invariant and evolutionary properties of the 
skew-symmetric differential forms. 

http://arXiv.org/pdf/math.GM/0401039

3. Cartan E., Les Systemes Differentials Exterieus ef Leurs Application 
Geometriques. -Paris, Hermann, 1945. 

4. Bott R., Tu L.~W., Differential Forms in Algebraic Topology. 
Springer, NY, 1982. 

5. Schutz B.~F., Geometrical Methods of Mathematical Physics. 
Cambridge University Press, Cambridge, 1982. 

6. Encyclopedia of Mathematics. -Moscow, Sov.~Encyc., 1979 (in Russian). 

7. Wheeler J.~A., Neutrino, Gravitation and Geometry. Bologna, 1960. 

8. Petrova L.~I., Identical and nonidentical relations. Nondegenerate  
and degenerate transformations. (Properties of the skew-symmetric  
differential forms).  

http://arXiv.org/pdf/math.GM/0404109 

9. Petrova L.~I., Conservation laws. Their role in evolutionary 
processes. (The method of skew-symmetric differential forms). 

http://arXiv.org/pdf/math-ph/0311008

10. Fock V.~A., Theory of space, time, and gravitation. -Moscow, 
Tech.~Theor.~Lit., 1955 (in Russian). 

11. Clark J.~F., Machesney ~M., The Dynamics of Real Gases. 
Butterworths, London, 1964. 

12. Dafermos C.~M. In "Nonlinear waves". Cornell University Press, 
Ithaca-London, 1974. 

13. Tolman R.~C., Relativity, Thermodynamics, and Cosmology. 
Clarendon Press, Oxford,  UK, 1969.

\end{document}